\documentclass[sigconf]{acmart}
\AtBeginDocument{%
  \providecommand\BibTeX{{%
    \normalfont B\kern-0.5em{\scshape i\kern-0.25em b}\kern-0.8em\TeX}}}

\usepackage{subcaption}

\begin{document}

\title{Algorithmic Ways of Seeing: Using Object Detection to Facilitate Art Exploration}


\author{Louie S{\o}s Meyer}
\orcid{0009-0008-7490-3049}
\authornote{Both authors contributed equally to this research.}
\affiliation{%
  \institution{IT University of Copenhagen}
  \streetaddress{Rued Langgaards Vej 7}
  \city{Copenhagen}
  \country{Denmark}
  \postcode{DK-2300}
}
\email{lomm@itu.dk}

\author{Johanne Engel Aaen}
\authornotemark[1]
\orcid{0009-0004-7395-5163}
\affiliation{%
  \institution{IT University of Copenhagen}
  \streetaddress{Rued Langgaards Vej 7}
  \city{Copenhagen}
  \country{Denmark}
  \postcode{DK-2300}
}
\email{jaae@itu.dk}

\author{Anitamalina Regitse Tranberg}
\orcid{0009-0007-5463-4307}
\affiliation{%
  \institution{IT University of Copenhagen}
  \streetaddress{Rued Langgaards Vej 7}
  \city{Copenhagen}
  \country{Denmark}
  \postcode{DK-2300}
}
\email{anitatranberg@gmail.com}

\author{Peter Kun}
\orcid{0000-0003-0778-7662}
\affiliation{%
  \institution{IT University of Copenhagen}
  \streetaddress{Rued Langgaards Vej 7}
  \city{Copenhagen}
  \country{Denmark}
  \postcode{DK-2300}
}
\email{peku@itu.dk}

\author{Matthias Freiberger}
\orcid{0000-0003-2101-6274}
\affiliation{%
  \institution{IT University of Copenhagen}
  \streetaddress{Rued Langgaards Vej 7}
  \city{Copenhagen}
  \country{Denmark}
  \postcode{DK-2300}
}
\email{ m.freiberger@gmail.com}

\author{Sebastian Risi}
\orcid{0000-0003-3607-8400}
\affiliation{%
  \institution{IT University of Copenhagen}
  \streetaddress{Rued Langgaards Vej 7}
  \city{Copenhagen}
  \country{Denmark}
  \postcode{DK-2300}
}
\email{sebr@itu.dk}

\author{Anders Sundnes L{\o}vlie}
\orcid{0000-0003-0484-4668}
\affiliation{%
  \institution{IT University of Copenhagen}
  \streetaddress{Rued Langgaards Vej 7}
  \city{Copenhagen}
  \country{Denmark}
  \postcode{DK-2300}
}
\email{asun@itu.dk}

\renewcommand{\shortauthors}{Meyer et al.}

\begin{abstract}
This Research through Design paper explores how object detection may be applied to a large digital art museum collection to facilitate new ways of encountering and experiencing art. We present the design and evaluation of an interactive application called SMKExplore, which allows users to explore a museum's digital collection of paintings by browsing through objects detected in the images, as a novel form of open-ended exploration. We provide three contributions. First, we show how an object detection pipeline can be integrated into a design process for visual exploration. Second, we present the design and development of an app that enables exploration of an art museum's collection. Third, we offer reflections on future possibilities for museums and HCI researchers to incorporate object detection techniques into the digitalization of museums. 
\end{abstract}


\begin{teaserfigure}
  \includegraphics[width=\textwidth]{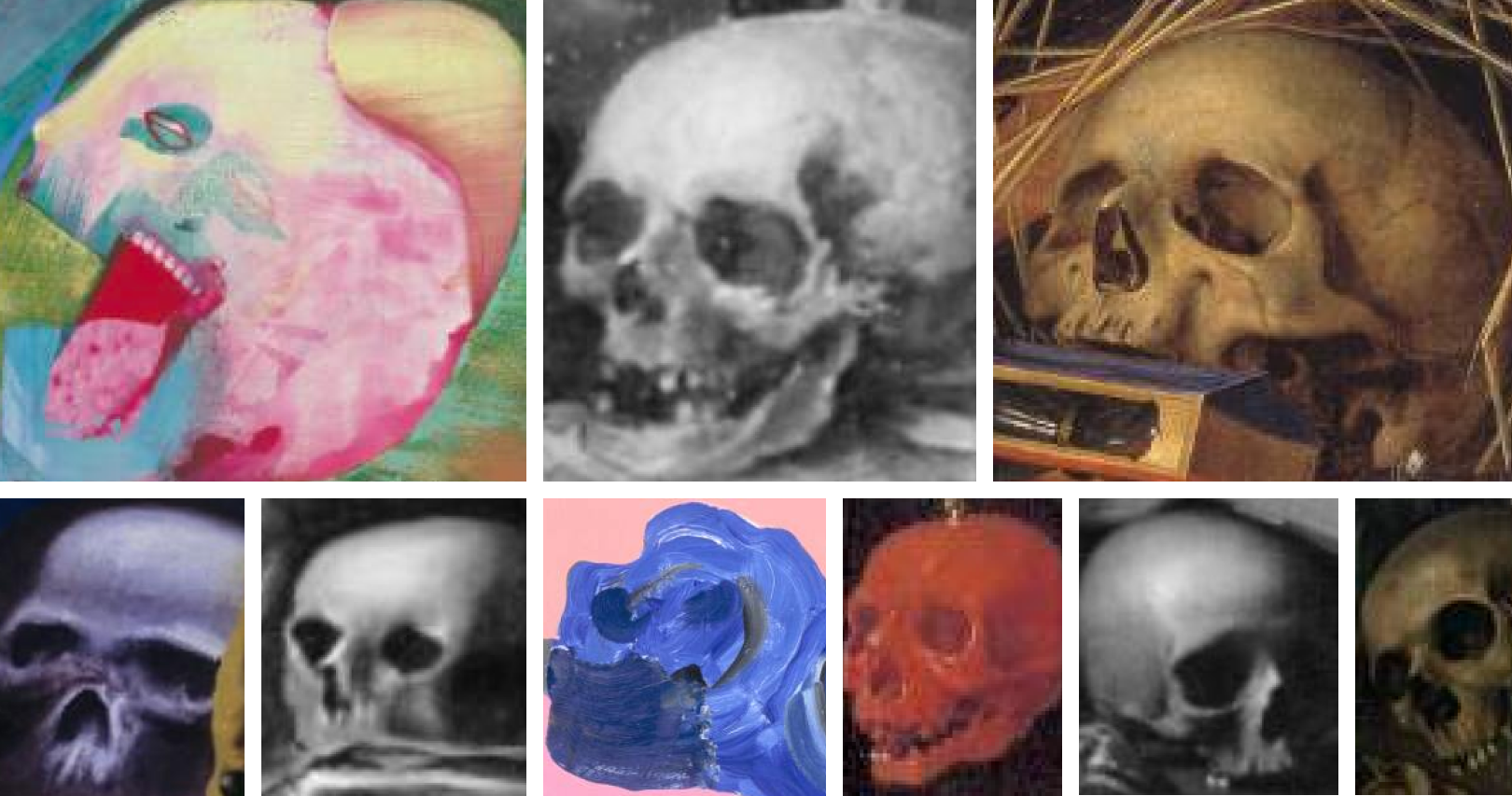}
  \caption{A selection of objects classified as ``skull'' in paintings from the collection of the National Gallery of Denmark.}
  \Description{A grid of images that have been cropped from digitized paintings. The images in the grid showcase objects that have been classified as skulls by an object detection model. Some of the images are realistic portrayal of human skulls while other images show motifs that are more abstract.}
  \label{fig:teaser}
\end{teaserfigure}

\received{14 September 2023}
\received[revised]{12 December 2023}
\received[accepted]{19 January 2024}

\maketitle
\section{Introduction}
Recent progress in computer vision has led to algorithms that are comparable to human performance on some tasks \cite{dodge2017study}. Previously, vision algorithms have been limited by what has been termed the ``cross-depiction'' problem \cite{hall_cross-depiction_2015}: While achieving impressive performance in detecting objects on photographic images, algorithms would struggle to detect the same objects in other depictions such as drawings, paintings, and other art styles. However, recent advances in machine learning algorithms trained on multimodal image-text datasets and approaches such as Contrastive Language-Image Pre-training (CLIP) \cite{radford2021_CLIP} and Grounded Language-Image Pre-Training (GLIP) \cite{Li_2022_GLIP} offer promising performance across visual domains. In this paper, we explore how these technologies may be applied to the large art collection of the National Gallery of Denmark (Danish abbreviation: SMK) to facilitate new ways of exploring and experiencing art. While techniques such as computer vision and, more broadly, Artificial Intelligence (AI) raise both legal and ethical concerns relating to authorship, bias, trust, and more \cite{crawford_excavating_2019,foka_critically}, these technologies also offer the potential to make art collections more accessible and to offer new ways of experiencing art.

As suggested by Lev Manovich, while visual art and aesthetics are traditionally experienced and studied by looking at individual images and artworks, computational analysis of images opens up the perspective of exploring large datasets, inviting us to shift our perspective from unique exemplars to \textit{``seeing one billion images''} and the patterns therein \cite{manovich_cultural_2020}. Within Human-Computer Interaction (HCI) such perspectives have been explored through the design of novel systems for visualization and exploratory search \cite{white_exploratory_2009, windhager_visualization, dork_information_2011}. Exploratory search is of particular importance for museums and art collections, because it allows   non-expert users to find pathways to explore and discover art that they don't know about and so wouldn't know how to search for in a traditional search interface - an issue that is strongly aligned with museums' mission to inspire and educate \cite{wray_pathways_2013, whitelaw2015generous, bjorneborn_serendipity}. 

Museums have long served as a productive environment for research and experiments in Human-Computer Interaction (HCI) \cite{hornecker_human-computer_2019}, and debates around the use of computer vision in museums have become increasingly prominent \cite{ciecko_ai_2020, villaespesa2021not}. The recent significant developments in object detection suggest that computer vision may be applied now to museum collections to make them searchable not just through metadata about the images but through the subject matter of the artworks - i.e. the objects appearing in the images. Presently, such search has been limited by the extent to which information about the objects has been manually entered into the collection metadata by museum curators - which is often far more sparse (if at all available) than the rich visual information in the images \cite{bacon_ai_2019, ciecko_ai_2020}. Thus, this paper presents a research through design \cite{zimmerman_research_2007} exploration of the following research question:

\begin{quote}
\textbf{RQ:} \textit{How can object detection be used to support exploration of an art museum's digital collection?}
\end{quote}

As this approach represents a novel application of object detection techniques, a significant part of the effort has been to implement an object detection workflow for extracting object data about the art collection. We present the design and evaluation of an interactive application, SMKExplore, which allows users to explore a museum's digital collection of art paintings by browsing through objects detected in the images, as a novel form of exploration. 

In this paper, we provide three contributions. First, we show how an object detection pipeline can be integrated into a design process for visual exploration. Second, we present the design and development of an app that enables exploration in the context of a museum collection. Third, we offer reflections on future possibilities for museums and HCI researchers to incorporate object detection techniques in digital museum collections.

\section{Related Work}
\label{Related Work}
\subsection{Object Detection and Artwork}
Along with rapid advances in machine learning, scholars have debated how designers may use machine learning as {\em design material} \cite{yang_ml_hard,gillies2016human,kuniavsky2017designing,benjamin2021machine,davis_drawing_apprentice,Dove2017}. In this paper, we focus on one particular application of machine learning: Object detection in art images.

With large numbers of home users gaining access to the world wide web in the early 2000s, large online image collections emerged as users shared their personal photographs and drawings. With the advent of these collections, research interest in automatic image annotation and image retrieval to make sense of them has strongly increased \cite{datta2008image}. Important early work includes probabilistic approaches to automatically match text and images \cite{barnard2003matching,blei2003modeling}, efficient scene matching techniques \cite{torralba2008small}, as well as web-based tools to crowd-source image-labelling at scale \cite{russell2008labelme}. 
Recent research directions on using large image-text datasets leverage deep learning \cite{lecun2015deep} to develop new approaches for various computer vision tasks.

While deep neural networks have shown great promise when trained to recognize different objects \cite{amit2020object}, image recognition systems are brittle and may struggle if images are slightly grainy or noisy and are vulnerable to manipulation \cite{dodge2017study, heaven_why_2019, su_one_2019}. Neural networks struggle even more to recognize objects depicted in different styles, such as drawings or paintings. This is known as the cross-depiction problem \cite{hall_cross-depiction_2015, westlake_detecting_2016}. The cross-depiction problem reveals a weakness in image recognition systems compared to human vision: While humans are versatile and can recognize even relatively minimal line drawings with ease, neural networks are highly specialized and perform poorly when confronted with an image style that is different from the data used in training \cite{boulton_artistic_2019}. One of the sources of this weakness might be that computer vision algorithms tend to be biased towards focusing on texture, unlike human vision, which tends to focus more on shape \cite{geirhos_imagenet-trained_2019}. However, texture bias can be reduced by modifying the training data (while using standard architectures) \cite{geirhos_imagenet-trained_2019}. Kadish and colleagues adapted the technique presented in \cite{geirhos_imagenet-trained_2019} and applied it to object detection on the artworks in the People-Art dataset, achieving a 10\% improvement in state of the art for this dataset  \cite{kadish_improving_2021}.

\begin{figure*}[h]
  \centering
  \begin{subfigure}[t]{0.48\textwidth}
    \centering
    \includegraphics[width=\linewidth]{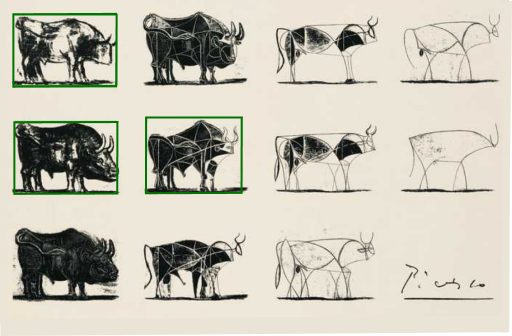}
    \caption{\textit{Le Taureau (The Bull)} by Pablo Picasso, analyzed in 2021 using the Fast R-CNN \cite{Girshick_2015_ICCV} algorithm pre-trained on COCO \cite{Lin_2014_ECCV}. The boxes drawn around three of the drawings indicate that these drawings are classified as ``cow'' by the algorithm.}
    \Description{A picture of 11 drawings of a bull. The drawings vary in their degree of realism; from quite realistic to a simple line drawing. Three of the most realistic drawings have bounding boxes around them, indicating they have been classified as ``cow'' by the object detection model.}
    \label{fig:subfiga}
  \end{subfigure}%
  \hfill
  \begin{subfigure}[t]{0.48\textwidth}
    \centering
    \includegraphics[width=\linewidth]{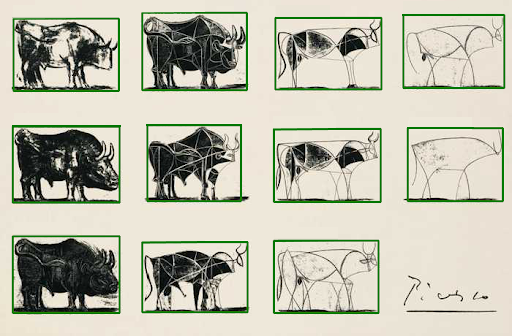}
    \caption{\textit{Le Taureau (The Bull)} by Pablo Picasso, analyzed in 2023 using GLIP \cite{Li_2022_GLIP}. The boxes drawn around each drawing indicate that they are all correctly classified as ``bull''.}
    \Description{A picture of 11 drawings of a bull. The drawings vary in their degree of realism; from quite realistic to a simple line drawing. All the drawings have bounding boxes around them, indicating they have all correctly been classified as ``bull'' by the object detection model.}
    \label{fig:subfigb}
  \end{subfigure}
  \caption{Comparison of image analysis results.}
  \label{fig:comparison}
\end{figure*}

Recently, great progress has been made with generative models such as Midjourney, DALL·E \cite {ramesh2022hierarchical_dalle,ramesh2021zero}, and Stable Diffusion \cite{Rombach_2022_CVPR_stablediffusion,sohl_2015_ICML}, which are enabled by contrastive training between images and text \cite{radford2021_CLIP,Li_2022_GLIP}. In this approach, a given image and its description text are mapped into a shared high-dimensional vector space. This approach has brought remarkable progress on zero-shot computer vision tasks for image-text retrieval, image captioning, or visual Q\&A \cite{pmlr-v162-li22n}. To illustrate the progress of these approaches on non-photographic content, consider Pablo Picasso's series of lithographs titled ``The Bull'', which demonstrates a range of depiction styles from a lifelike drawing to ever more abstract styles. Fig. \ref{fig:subfiga} shows the result of a test the authors of this paper ran in 2021, exploring how well a state-of-the-art image recognition algorithm (Fast R-CNN \cite{Girshick_2015_ICCV}, pre-trained on Common Objects in Context (COCO)) \cite{Lin_2014_ECCV} could classify these drawings. At this time, the algorithm could only identify three drawings showing a ``cow'' (as the COCO labelset had no separate category for bull). In 2023, the authors tested the Grounded Language-Image Pre-Training (GLIP) \cite{Li_2022_GLIP} algorithm on the same image, correctly classifying even the most abstract drawings as bulls (see Fig. \ref{fig:subfigb}). This illustrates that these new algorithms have enabled a great leap in cross-depiction object detection, now making it feasible to use these technologies on art collections with a hitherto unmatched degree of precision.

\subsection{AI and Museums}
Although recently a surge of artists have been working with generative AI systems to create art, sometimes called ``AI Art'' \cite{audry2021art,bogost_ai-art_2019,cetinic_creating_art_ai,zylinska_ai_2020,oppenlander,boden,epstein_science_art_ai}, in this paper we focus on the use of computer vision technology for \textit{Experience Design} in an art museum context, rather than creating new forms of art.

There is a long history of HCI research conducted in collaboration with museums, both using the museum context as a testbed for new technologies, as well as drawing insights from the interplay of technology with art and heritage collections \cite{hornecker_human-computer_2019,waern_hybrid_2022}. Museums around the world have spent much efforts in digitizing their collections, both as a way to secure the contents of the collections for the future and also to increase public access -- particularly to the vast number of artifacts in archives that cannot be exhibited physically in the museums. It is commonly suggested, that large European museums usually have a very small share of their collections on display in their physical exhibition spaces, both due to limitations in space and other capacity limitations, as well as many artworks being too brittle to exhibit. In the case of the National Gallery, only 0.7\% of the collection is on display \cite{noauthor_smk_2018}. Making digital reproductions of the collections available online allows for the public to access the entirety of the collection at any point in time. This digitalization agenda in the museum and cultural heritage sector has also fostered the implementation of AI techniques for collection management, audience predictions, art authentication, and more \cite{ciecko2017examining,ciecko_ai_2020,villaespesa2021not}.

Museum collections often contain vast amounts of data that can be used for many purposes, including categorizing and collection management, as well as a means to investigate patterns of (dis-)similarities between artists, cultures, time periods, or even across an entire collection  \cite{Giannini2019, meinecke2022labeling}. Describing and allocating metadata to museum and cultural heritage items is an essential yet labor-intensive task, which often requires highly specialized domain expertise from museum curators and researchers \cite{meinecke2022labeling}.

There is a growing interest in applying computer vision in museums \cite{ciecko_ai_2020, villaespesa2019museum,villaespesa2021not}. Museums commonly have experimented with computer vision techniques to enrich their metadata \cite{tallon2018creating, smith_smks_2019, stack2020computervision}. Such algorithmic enrichment may be helpful both for museum researchers who frequently need to search their collection, as well as for non-expert audiences. For example, the Harvard Art Museum invites online users to investigate how computers see and process images of artwork, comparing metadata generated by humans to metadata generated by AI technologies developed by Amazon, Clarifai, Imagga, Google, and Microsoft \cite{HavardArtMus}. Another example is the Princeton Art Museum, which has experimented with computer vision for various purposes, including detecting visual similarities in Chinese paintings from different eras \cite{villaespesa2021not}. Museums have also explored using computer vision techniques in experiences for visitors to the physical museum exhibition, for instance through mobile apps which use image recognition to allow visitors to point their camera at an artwork and receive information about it \cite{macdonough_smartify_2018,lovlie_designing_2021,theodosiou_artwork_recognition_2022}.

Although object detection shows great promise for museums, there are potential pitfalls concerning bias. Museums hold widespread artworks in their collections, among them pieces representing controversial, challenging, and painful parts of history and contemporary society \cite{villaespesa2021not}. Research has found that computer vision is limited by bias, specifically in terms of cultural and gender biases \cite{ciecko_ai_2020, villaespesa2021not}. Ciecko and colleagues \cite{ciecko_ai_2020} question whether museums should withhold some artworks from being classified by computer vision algorithms, highlighting colonization, slavery, and genocide as particularly challenging topics. The art collection used in the study at hand is a broadly themed collection of historical and modern art and does not focus specifically on such challenging topics. However, it is hard to predict where bias may appear when analysing a large and varied collection of artworks. In this study, the results of the computer vision analysis were only presented to a small group of test participants, thus reducing the risk of unintended offense or controversy. If the design presented here should be made available to the public at large, the museum would need to carefully consider the risks associated with bias and possible mitigation strategies.

\subsection{Exploratory Search in Digital Collections}
The metadata of an art museum's digital collection is a complex information space, as these collections are constructed and used by different professionals performing various complex tasks that go beyond ``variants on a search box'' \cite{ruecker2016visual}. As an alternative to restricted and result-oriented keyword-based search, \textit{exploratory search} emerged in HCI to support open-ended, interactive, and evolving processes as a strategy for information seeking \cite{marchionini2006exploratory, KULES2008463, Yee_faceted_2003}. While it lacks a rigid definition, its essence revolves around the journey of searching rather than the user arriving at precisely defined outcomes \cite{palagi2017survey,soufan2022searching}. Unlike the linear trajectory of keyword searches, exploratory search is iterative; it fosters a dynamic evolution of user needs as they garner new insights \cite{russell2012designing, navrat2012cognitive, white_exploratory_2009}. Exploratory search encourages leisurely browsing, inviting users on unexpected voyages through data \cite{dork_information_2011}. Ultimately, its goal leans towards an engaging user experience rather than just efficiently returning search results \cite{wilson2010keyword}.

The visual nature of cultural heritage collections has afforded visualization strategies to enable experts and non-experts to interact with such datasets, also motivated by going beyond keyword-based searches \cite{windhager_visualization}. A specific type of interactive visualization of cultural heritage collections is described as \textit{generous interfaces} which are rich, browsable interfaces that reveal the scale and complexity of digital heritage collections \cite{whitelaw2015generous}. Such interfaces have shown that multiple entry points to a collection and navigating via multiple paths allow rich opportunities for exploration and discovery \cite{thudt2012bohemian, Junginger_Ostendorf_Avila2020,damiano2019investigating}. 

Many museum visitors might be unaware of the collection's contents, potential attractions, or even their own interests and objectives during a visit \cite{falk_museum_2012, mancas_hypersocial_2009}. For this reason scholars working with information visualization and design relating to museums and libraries have often been interested in facilitating serendipitous discovery \cite{windhager_visualization, thudt2012bohemian, damiano2019investigating}, meaning chance encounters with items of interest \cite{bjorneborn_serendipity, dork_information_2011}. Cole and colleagues \cite{Cole_Dau_Ducrou_Eklund_Wray_2019} present several approaches to facilitating exploratory search and serendipitous discovery using techniques such as similarity search and formal concept analysis (see also \cite{wray_pathways}). Thus, the concepts of exploratory search and generous interfaces offer a helpful perspective in designing for exploration of art collections. This paper contributes by investigating how to apply object detection to create a novel interface for exploration of the collection.

\section{Approach and Methodology}
The current project is a research through design \cite{zimmerman_research_2007} exploration based on an interdisciplinary collaboration between scholars in machine learning and human-computer interaction (HCI). The overarching goals for the project are to use the interdisciplinary collaboration to break new ground both for computer vision - applied to art images of different depiction types and styles - as well as for human-computer interaction, in particular regarding the use of machine learning as a design material applied to experiences with visual art. One of the authors of this paper has extensive experience with HCI research in art museums, but none of the authors have specific domain expertise in art analysis.

Thus, the project's starting point was a technical exploration of the state of the art in object detection on art images, carried out by two of the authors during the second half of 2022. This led to the setup described in section \ref{sec:ObjectDetection} below, using GLIP to annotate all 6,750 paintings in the National Gallery's digital collection with object labels.

Once this material had been established we began a design exploration aiming to create an interactive application allowing a general museum audience to browse and experience the collection in novel ways. This was done through an iterative process involving stakeholders at the National Gallery and their audiences from February to August 2023. This process also revealed a need to revise and iterate on the object annotations, as described below in section \ref{sec:custom-labels}.

The resulting interactive application was evaluated with test users recruited on-site at the National Gallery 11-12 Aug 2023, as described in section \ref{Evaluation} below.

\section{Object detection}
\label{sec:ObjectDetection}
Since no ground truth is available on the National Gallery's collection, our object detection approach has relied on pre-trained models. With COCO \cite{Lin_2014_ECCV} being the most prominent object detection dataset available, many approaches applied to artwork are trained or pre-trained on it; for instance, see \cite{kadish_improving_2021,springstein_semi_2022}. Designed to mirror modern-day visual environments, the COCO dataset emphasizes contemporary objects, rendering its classes heavily limited for the nuanced motifs or historical themes of traditional art paintings. For this reason, we decided to utilize a contrastively pre-trained model with the possibility to customize the labels of detected objects in a way suitable in the context of artwork. 

Our approach involves defining a set of labels (up to 120), which is then presented as a single string of words separated by full stops to the GLIP model \cite{Li_2022_GLIP}. The pre-trained GLIP model then represents each class label as a vector. This vector representation of an object label can then be compared to similar representations extracted from image patches. The model matches the image patch and object label vectors most similar to each other and generates labeled bounding boxes in the image based on these similarities.
To evaluate the principal suitability of our approach for art images, we have applied it to the People-Art \cite{westlake_detecting_2016} test set, where we achieved an average precision (AP) of $0.56$ (label = ``person'', confidence cutoff = $0.25$, intersection over union $0.5-0.95$). In comparison, recent work \cite{kadish_improving_2021,springstein_semi_2022} reported APs of $0.36$ and $0.44$, respectively. We therefore concluded that our approach is suited for object detection in the National Gallery's collection. 

\subsection{Defining Custom Labels}
\label{sec:custom-labels}
\begin{figure*}[]
  \centering
  \begin{subfigure}[t]{0.5\textwidth}
  \centering
    \includegraphics[width=\linewidth]{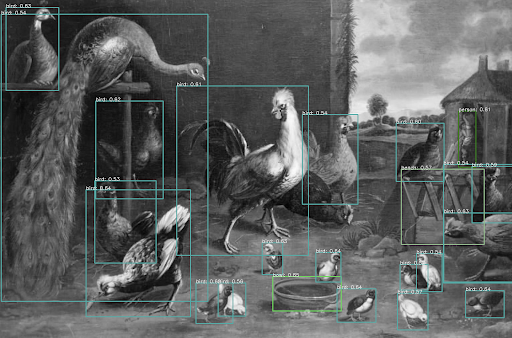}
    \caption{The painting ``Between hens, roosters and chickens, a few peacocks can be seen'' (1749 – 1848) by Unknown Artist from the collection of the National Gallery of Denmark, analyzed by GLIP. Even in a crowded composition with many overlapping objects the algorithm is able to correctly label a large number of objects.}
  \Description{A painting of various birds with bounding boxes around the objects the model has detected.}
  \label{fig:cluttered}
  \end{subfigure}%
  \hfill
  \begin{subfigure}[t]{0.4\textwidth}
  \centering
    \includegraphics[width=\linewidth]{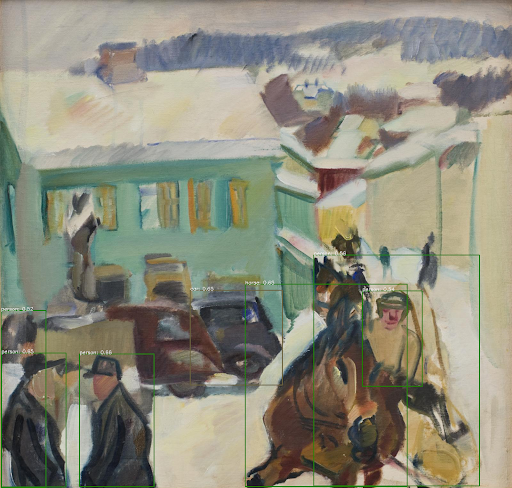}
    \caption{The painting ``Lillehammer. February. Timber Sledges'' (1937) by Harald Leth from the collection of the National Gallery of Denmark, analyzed by GLIP. The algorithm can detect objects even in a relatively non-realistic painting style.}
  \Description{A non-realistic painting of a snowy town landscape with bounding boxes around the detected objects.}
  \label{fig:abstract}
  \hspace{5em}
  \end{subfigure}

  \begin{subfigure}[t]{0.5\textwidth}
  \centering
    \includegraphics[width=\linewidth]{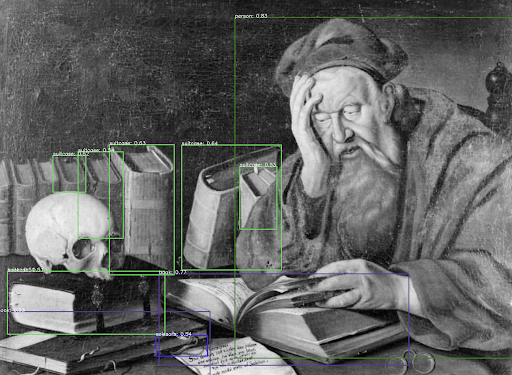}
   \caption{The painting ``Saint Jerome'' (16th century) by Unknown Dutch Artist from the collection of the National Gallery of Denmark, analyzed by GLIP. The books are mislabelled as ``suitcase''.}
  \Description{A realistic painting with bounding boxes around several books, incorrectly labeled as ``suitcase''.}
  \label{fig:suitcases}
  \end{subfigure}
  \hfill
  \begin{subfigure}[t]{0.4\textwidth}
  \centering
    \includegraphics[width=\linewidth]{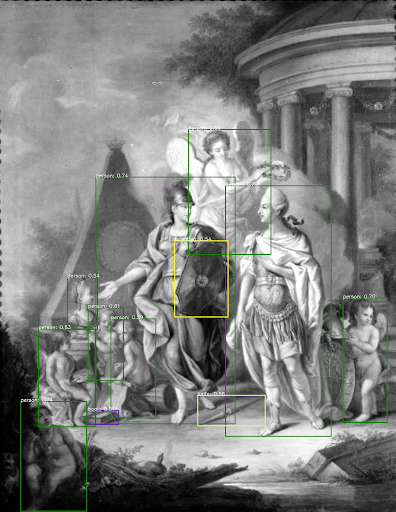}
    \caption{The painting ``Christian VII as the protector of art and science. Allegory'' (1770) by Georg Mathias Fuchs from the collection of the National Gallery of Denmark, analyzed by GLIP. The shield of the warrior figure is labelled as ``handbag''.}
  \Description{A painting in classical style showing a bounding box around a shield, labeled as ``handbag''.}
  \label{fig:handbag}
  \end{subfigure}
  \caption{Comparison of image analysis results.}
  \label{fig:classification-examples}
\end{figure*}

In order to minimize errors, we built a dataset consisting only of digitized artworks from the museum collection that were labeled as ``painting'' in the collection metadata, resulting in a set of 6,750 artworks. Initially, we applied the approach described above, using the 80 object categories in COCO. This gave impressive results: 
The system could recognize a wide range of objects even in crowded scenes (Fig. \ref{fig:cluttered}) and with unclear depiction styles (Fig. \ref{fig:abstract}). However, the system appeared to have a bias towards modern object categories: for instance, mislabelling some old books as \textit{suitcases} (Fig. \ref{fig:suitcases}) or the shield of a female warrior figure as a \textit{handbag} - in the latter case perhaps also revealing a gender bias.

In order to mitigate this problem we pivoted to IconClass, a comprehensive index for classification of objects depicted in art images \cite{couprie_iconclass_1983}. We needed to construct a custom set of labels, as IconClass contains over 28,000 unique concepts, whereas our system could only accept a maximum of 120 labels. We iteratively explored categories and labels from IconClass and observed the frequency of such objects in a random subset of the National Gallery's paintings. As the first iteration of object detection with the COCO labels had shown a strong dominance of the label \textit{Person} (44\% of all the detected objects), we prioritized including various labels relating to people and clothing. However, we omitted labels for small details such as mouth and eyes, as we expected such objects would most often be too small to crop in high-resolution and, therefore, difficult to apply in our user interface. Furthermore, we also included several labels relating to themes we observed occurring often in the artworks, such as religious themes, architecture, food items, musical instruments, furniture, weapons, vehicles, and nature. This process resulted in a list of 120 labels, as seen in Table \ref{tab:category-labels}. 

\begin{table*}
  \caption{Custom set of labels that were used for object detection with the GLIP model.}
  \label{tab:category-labels}
  \begin{tabular}{ll}
    \toprule
    Category & Labels \\
    \midrule
     Animal & Bird, Butterfly, Cat, Chicken, Cow, Dog, Donkey, Fish, Horse, Insect, Mouse, Rabbit, Reptile, Sheep\\
     Architecture & Bridge, Castle, Church, Door, House, Mill, Pillar, Staircase, Window\\
    Christianity & Angel, Cross, Devil, God, Jesus Christ, Saint, Virgin Mary\\
     Clothing & Bag, Belt, Cane, Crown, Dress, Gloves, Hat, Jewellery, Mask, Shoes, Tie, Umbrella\\
     Food & Apple, Banana, Bread, Cheese, Grapes, Lobster, Orange, Pineapple, Vegetable, Watermelon, Wine\\
    Furniture & Bathtub, Bed, Chair, Easel, Sofa, Table\\
     Human & Baby, Child, Face, Hand, Man, Woman\\
   Instrument & Drum, Flute, Guitar, Harp, Piano, Violin\\
     Interior & Bird Cage, Book, Bottle, Bow, Cup, Drapery, Flag, Globe, Lamp, Mirror, Paper, Vase \\
     Nature & Bush, Cloud, Fire, Flower, Lake, Lightning, Moon, Mountain, Plant, Rock, Sea, Sky, Sun, Tree\\
     Occultism & Demon, Ghost, Skeleton, Skull, Star\\
    Vehicle & Airplane, Bicycle, Boat, Car, Carriage, Ship, Train, Wheel\\
     Weaponry &  Armor, Arrow, Bow, Firearm, Hammer, Helmet, Rope, Shield, Spear, Sword,\\
    \bottomrule
  \end{tabular}
\end{table*}

\subsection{Selecting a Subset}
Based on 6,750 paintings from the museum collection and the aforementioned 120 labels, a total of 109,145 objects were detected on 6,477 of the paintings. Analyzing the data revealed a skewed distribution of objects with 4 categories (\textit{Human}, \textit{Nature}, \textit{Architecture}, and \textit{Clothing}) representing more than 70\% of the total objects detected. Similarly a high variance existed in the number of objects detected per label: the object \textit{Man} had an instance count of 5,975 whereas \textit{Bird Cage} only had a total count of 5. 
Due to technical constraints, we needed to reduce the dataset to one-tenth of the total data. In order to get a more uniform representation of objects, we defined the subset by retrieving up to 100 object instances per label, selected based on highest confidence level. This resulted in a dataset consisting of 10,775 objects detected on 3,906 of the paintings from the collection. 

\begin{figure*}[h]
  \centering
  \includegraphics[width=0.84\textwidth]{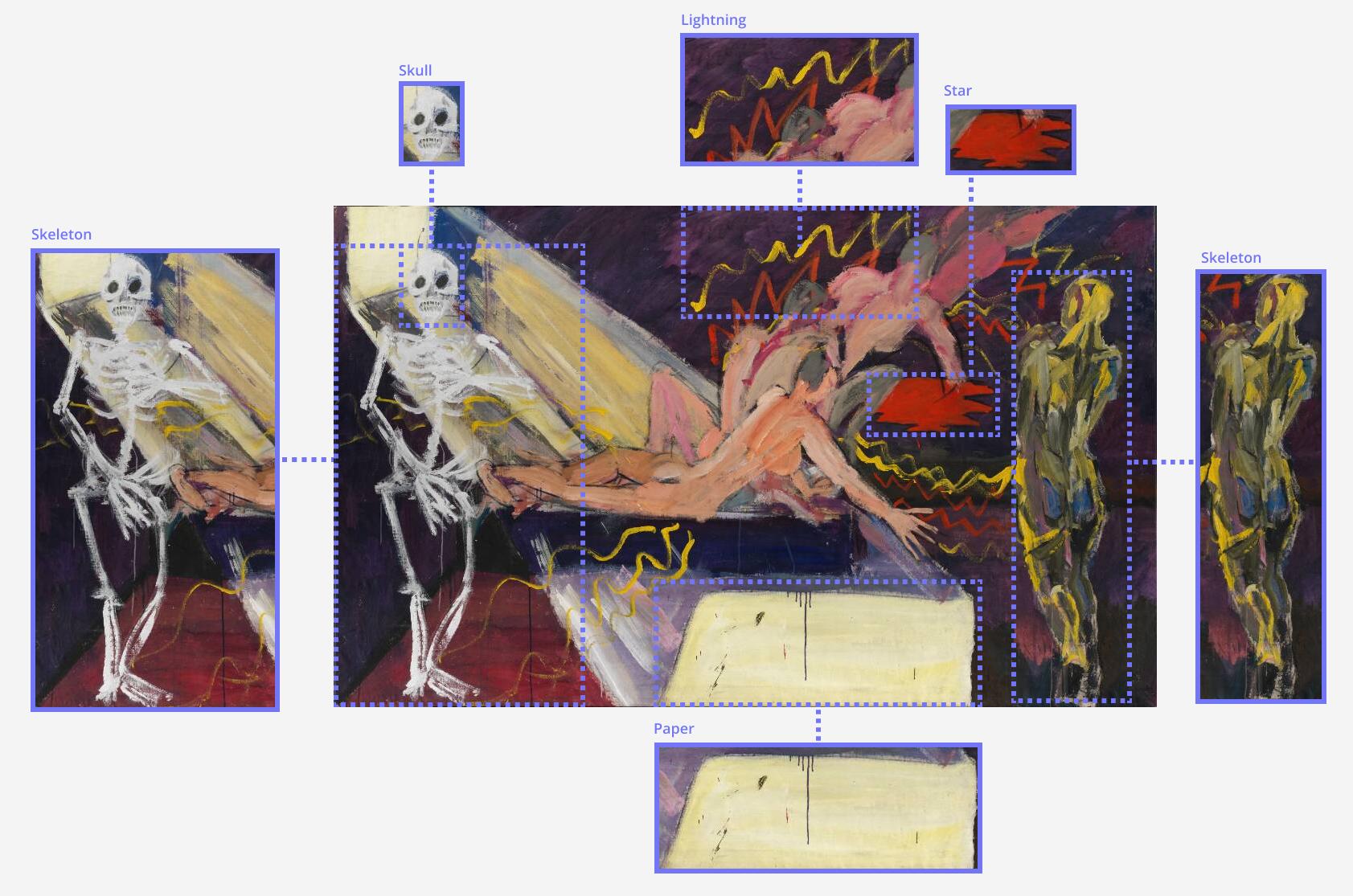}
  \caption{Objects detected in \textit{``The King and Queen Surrounded by Swift Nudes''} by Inge Ellegaard (1982).}
  \Description{Painting with five bounding boxes around detected objects respectively classified as a skeleton, skull, lightning, star, and a paper. The objects are also shown outside of the painting, to exemplify how they have been individually cropped from the painting.}
  \label{fig:cropped-images}
\end{figure*}

Finally we developed a script that cropped the detected objects from images of the original paintings, leaving us with an image collection of the singular objects as illustrated in Figure \ref{fig:cropped-images}. These individual object images were used as a key component throughout the design and development of the interactive application.

\section{SMKExplore}
\label{design-section}
Working with the data as described above gave us valuable insights that helped inform our design process. In combination with the results from the object detection, we drew inspiration from existing literature on designing for exploration (as summarized in Section \ref{Related Work}) to create SMKExplore: A web application that allows users to browse and explore a digital art collection through the objects detected in the paintings, as well as to use the objects to create new images using a generative image algorithm. In the following subsection we first present the design process and insights leading up to the final design, which is presented in the subsequent subsection.

\subsection{Design Process}
The design process leading to the design of SMKExplore ran from January-August 2023. The design and development was carried out by the three first authors of this paper, and was structured as a combination of UX Design and Agile Software Development using SCRUM, thus combining a total of five design sprints and software development sprints based on the model presented in \cite{hartson2019}.

From the outset, our aim was to explore how object detection data could be used to facilitate exploration of a digital museum collection, using the object data to create new entry points and alternative ways of browsing. The data processing described in Section \ref{sec:ObjectDetection} was conducted prior to the design process. Exploring the data helped us frame the design space and guide the process. 

During the initial phases we investigated opportunities and qualities inspired by techniques from interaction-driven design as presented by \cite{maeng2012interaction}. This led us to design a preliminary concept that fostered immersive interactions with the objects in a 3D gallery which we envisioned implemented in WebVR. However, we had concerns about complications regarding usability and motion sickness (cf. \cite{chong2021virtual}) as well as technical feasibility, and chose instead to develop a simpler 2D concept.

Based on insights from the research literature, we established a set of design principles for our system. First, as suggested by \cite{windhager_visualization} and \cite{dork_information_2011} we emphasized finding a balance between overview of the data as well as opportunities to explore information in detail. This led us to establish a clear information hierarchy that allows the user to gain an overview of the collection, while we also designed pathways to detailed information on each artwork. Dörk and colleagues \cite{dork_information_2011} furthermore inspired us to use \textit{visual cues} to design navigational paths and enhance the possibility of serendipitous discoveries. Inspired by \cite{windhager_visualization, damiano2019investigating} we additionally decided to enable users to save objects that caught their interest, as a means to revisit parts of the collection they enjoyed.

Furthermore, inspired by \cite{wilson2010keyword, marchionini2006exploratory} we decided to cluster the objects based on similarity and to provide possibilities to filter the data as a means to create overview and support various information-seeking strategies, such as comparing, combining and evaluating. Insights from \cite{marchionini2006exploratory, thudt2012bohemian, whitelaw2015generous} led us to design for accessing the collection through multiple entry points
and navigating via multiple paths (cf. \cite{dork_information_2011}), in order to enhance the sense of free exploration and varied forms of interaction with the items of the collection.

In addition to this, past research \cite{lopatovska2013exploring, hinrichs2008emdialog} has highlighted the benefits of using playful elements to advance and encourage non-expert user engagement and maintain user attention. Thus, we decided to include a playful element in the application in the form of an interactive canvas where users, with the help of generative AI, can create their own art with objects of personal interest from the collection.

With the aim to establish a clear connection between the application, SMKExplore, and the National Gallery, we defined the visual identity with inspiration from the museum's website. In particular, we drew upon the color scheme, font types, square frames and the layout of the Painting Screen (Fig. \ref{fig:art-detector-screens}).

The design was revised and implemented through five iterations (sprints), informed in part by usability testing and in part by technical aspects, until an initial prototype was tested on users in May 2023. Based on findings and feedback from this test a last revision of the design was conducted in the beginning of August 2023.

\subsection{Final Design}

\begin{figure*}[h]
  \centering
  \vspace{1em}
  \includegraphics[width=0.92\textwidth]{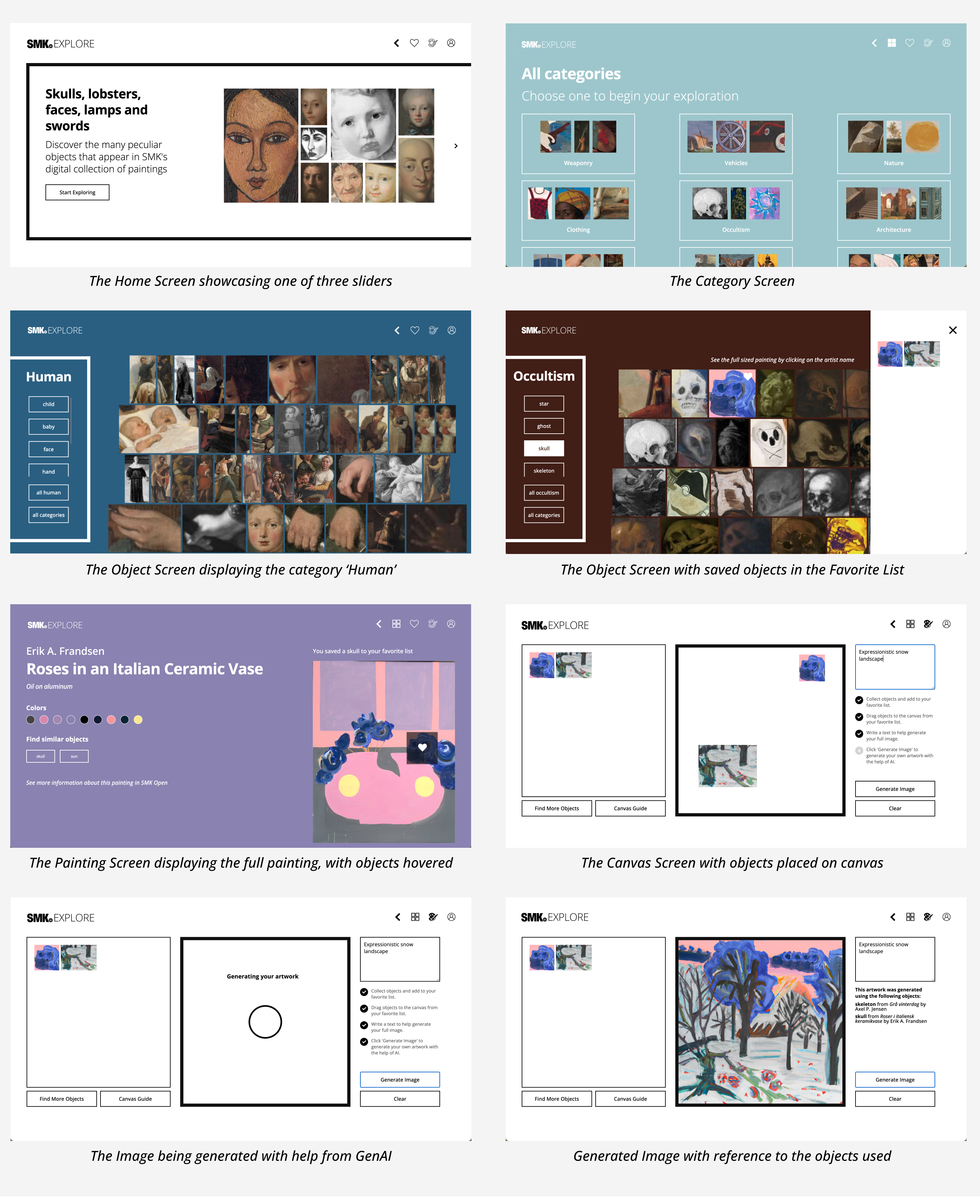}
  \caption{Screens from SMKExplore.}
  \Description{8 screens from the application SMKExplore, showcasing the layout and overall functionality of the application.}
  \label{fig:art-detector-screens}
\end{figure*}

In the following, we present the version of SMKExplore (Fig. \ref{fig:art-detector-screens}) that was used during the evaluation as reported in section \ref{Evaluation}.

SMKExplore allows \textit{``bottom-up search''}, where the user encounters the digital collection moving from details (i.e., objects) to the full-sized painting they appear in originally. The design concept focuses on navigating the collection based on thematic interest, shunning more traditional goal-oriented search. The aim is to allow users to compare depictions of similar objects in different paintings across time periods, styles, etc., and to help users discover artworks and details they otherwise might not have noticed.

When the user enters the application, they are met by the Home Screen, which showcases a slider with three examples of objects that have been detected in the collection. By clicking the ``Start Exploring'' button, they are led to a screen displaying the 13 categories defined for the objects in the data processing (see Table \ref{tab:category-labels}). The Category Screen constitutes the first level of an information hierarchy, in which the objects are presented only as a high-order category.

Once the user chooses a category they are directed to the Object Screen, where all objects within that particular category are displayed. Users may choose to filter the category further by selecting a label, thus being presented only with images of one object label, for instance skulls in the category Occultism.

Clicking on an object leads the user to the Painting Screen, which presents the entire painting on which the object appears. Alongside the painting more detailed metadata are provided, such as title, artist, technique, production year, and color palette. Other detected objects on the painting are also displayed, making it possible to navigate to other types of objects. Detected objects can also be discovered by hovering over the painting itself.

Users can save objects that catch their interest to a list of favorites. The list of favorites provides  users with an opportunity to revisit these parts of the collection later on. The saved objects can also be used to create new imagery using the interactive canvas. The canvas utilizes the outpainting function of OpenAI's DALL·E 2 API, which generates an image based on visual input(s) and a text prompt. The user may place objects on the canvas, resizing them as needed and leaving a generous amount of white space. Once they are satisfied with the composition they type in a text prompt describing, for instance, the desired image's style or theme. When the image is generated, the user has the possibility of creating a new image using the same or other objects, searching the collection further, or comparing their image to the original paintings by navigating through the list displaying the objects that were used on the canvas.

\section{Evaluation}
\label{Evaluation}
SMKExplore was evaluated on-site at the National Gallery 11-12 Aug 2023 during museum opening hours. Two authors were placed in the museum's foyer, inviting visitors to participate in a short user test. All test participants were visitors we encountered at the museum and were unknown to us beforehand. In total 22 participants interacted with the application and were interviewed, aged 18 to 59 (median: 26), 9 males and 13 females. The participants represented 14 nationalities across Europe, North America, Asia and Oceania. 

\subsection{Procedure}
Before interacting with SMKExplore, the content and functionality were briefly explained to the participants. Subsequently, they were instructed to explore the application freely, without time constraints. Towards the end of the session the participants were asked to generate an image on the Canvas Screen. The interactions with the application were documented using system logs as well as screen recordings.

A semi-structured interview was conducted immediately following the participants' interaction with the application. The interviews focused on the participants' experience of utilizing detected objects as the primary visual entry point to the collection. The participants were also asked about their general thoughts on using AI in art contexts and whether they noticed or reflected upon mislabelled objects. The interviews were recorded using an audio recorder. All participants were informed about the data collected and signed an information statement in accordance with the university's policies and the General Data Protection Regulation (GDPR).

The following analysis is based on findings from the system logs, interviews, and observations during the tests. Screen recordings have been used to supplement and clarify some details. The system logs were analyzed through descriptive statistics. The interviews were analyzed through thematic analysis, following the phases and guidelines presented by Braun and Clarke \cite{braunclarke_thematic_2006}. Additionally, following the guidelines by McDonald et al. \cite{10.1145/3359174}, consistency and validity of qualitative results was ensured through agreement among the authors by collaboratively developing the coding schemes through iterative discussions.

The initial phase of the thematic analysis was conducted by the first and second authors, who both had been part of the design team. They initially familiarized themselves with the data by transcribing and iteratively reading the interviews. From the transcripts, one author generated the initial codes for all interviews utilising the software ATLAS.ti. This amounted to a total of 158 distinctive codes related to our research question. These codes were assessed and revised by the second author. Subsequently, 16 groups of codes were established combining patterns such as ``questioning own interpretation'', ``noticing details through objects'' and ``surprised by personal interest''. In the following phases two additional authors, who did not take part in the design process nor the interviews, took part in the analysis and discussion to broaden the perspective. Through this process we found several overlaps in the 16 groups, which led us to narrow down to six potential themes touching upon object representation, attention to detail, interest driven search and discoveries, mislabelling and interpretation, and lastly contextualising the objects. Through an additional, conclusive, phase of analysis the themes and underlying patterns were reevaluated and the final six themes were established. These will be unfolded in the following.

\subsection{Overall Experience}
In general the participants became immersed rather quickly in browsing through the objects, looking concentrated throughout the process. When interacting with the canvas, they became more talkative towards the end of their session and seemed both entertained and surprised by the resulting image they generated. They spent between 4 and 15 minutes with the application (median: 8 minutes). The majority wanted to continue their exploration or said they could imagine themselves trying it again.

Generally, the participants interacted intuitively with the different features. They initially navigated from the Home Screen to the Category Screen and onward to the Object Screen. On their first visit to the Category Screen, they quickly (on average 10 seconds) found a category of interest to investigate further. Only 3 participants needed guidance on how to save an object to their favorite list. In addition, 4 participants said the functionality of the canvas could have been more apparent to them, while 3 participants said they could have used more tips on how it works. These issues primarily concerned confusion about how to resize objects on the canvas. 

20 of the 22 test participants said they enjoyed using the application and described the overall experience with words such as ``fun'', ``interesting'', ``intuitive'', ``enjoyable'', and ``ludic''. 2 participants had somewhat more mixed feelings.

\subsection{Representation of Objects}
The participants generally spent the most time on the Object Screen (see Table \ref{tab:avg-time-screens}), which shows all the detected objects for a specific label or category. During the interviews, several participants shared that exploring the collection through objects made them reflect on the different depictions of these objects and the wide range of motifs represented in the collection. \textit{``It was nice to see bikes from different paintings [...] I have never thought about looking at paintings and being like, oh, this is a bike here, and there is also a bike there''} (P1). Several participants mentioned the wide variety of object types as an element of surprise to them: \textit{``Wow, there are many of these objects I have never noticed in many of the artworks before''} (P7).

\begin{table}[h!]
  \caption{The average time spent on each screen of the application.}
  \Description{A table describing the average time spent on each screen of the application}
  \label{tab:avg-time-screens}
  \begin{tabular}{ll}
    \toprule
    Screen & Average time spent\\
    \midrule
    Object Screen & 3 Minutes \& 22 Seconds\\
    Canvas Screen & 3 Minutes \& 4 Seconds\\
    Painting Screen & 1 Minute \& 27 Seconds\\
    Category Screen & 27 Seconds\\
    Home Screen & 14 Seconds\\
  \bottomrule
\end{tabular}
\end{table}

In addition to the rich variation of objects, the participants commented on the effect of seeing the different depictions of these objects side by side on the Object Screen: \textit{``Many motifs are reappearing. It makes sense, but when you see it like this, it is wild''} (P5). The participants shared how distinct representations of the same objects made them reflect on different styles of painting through time. \textit{``It shows how different artists from different parts of the world, during different times have treated that object. Say, an apple would be very different in the Renaissance than today''} (P16).

\subsection{Focusing on Details}
When asked to describe their experience of accessing the collection primarily from the objects as opposed to the entire painting, several participants emphasized that it offered them a perspective on the artwork that made them notice things they would usually disregard. Removing the objects from their original context also made many aware of the complexity that goes into a painting, which they might not have discovered otherwise. In addition to this, several of the participants also expressed that experiencing the collection in this manner made them pay attention to what they were seeing and inspired them to look at the details more: \textit{``I think you just become more thoughtful of what actually is happening in a painting like this and what is present''} (P1).

Some also suggested that focusing on details can serve as an interesting new way to discover the entire paintings, as browsing the objects made them aware of artworks that they had not noticed when going through the exhibition: \textit{``[...] by going through details that maybe struck me, I also had the chance to pay attention to paintings that maybe I disregarded in the exhibition''} (P3).
 
While most participants enjoyed or found it interesting to experience the collection through the objects, some also expressed lacking the context of the objects as problematic or something they did not enjoy. Particularly, worries about losing the entire vision of the artist, were mentioned by those who would rather see the entire painting up front: \textit{``I like the whole vision that the artist had rather than just a small piece of it that somebody else had decided I would look at''} (P4).

\subsection{Interests and Discoveries}
A recurring pattern in the participants' interview answers was the ability to explore the museum collection based on their personal interests. They shared reflections on how this influenced their navigation in the application and that they discovered patterns in what caught their attention: \textit{``I learned what I am interested in when I look at art''} (P19). Several participants stated they had found new and unexpected artworks by pursuing their interest in particular objects: \textit{``I didn't think I would be interested in a painting of cows, but that was very surprising and interesting''} (P16). 

\begin{figure}[h]
  \includegraphics[width=0.46\textwidth]{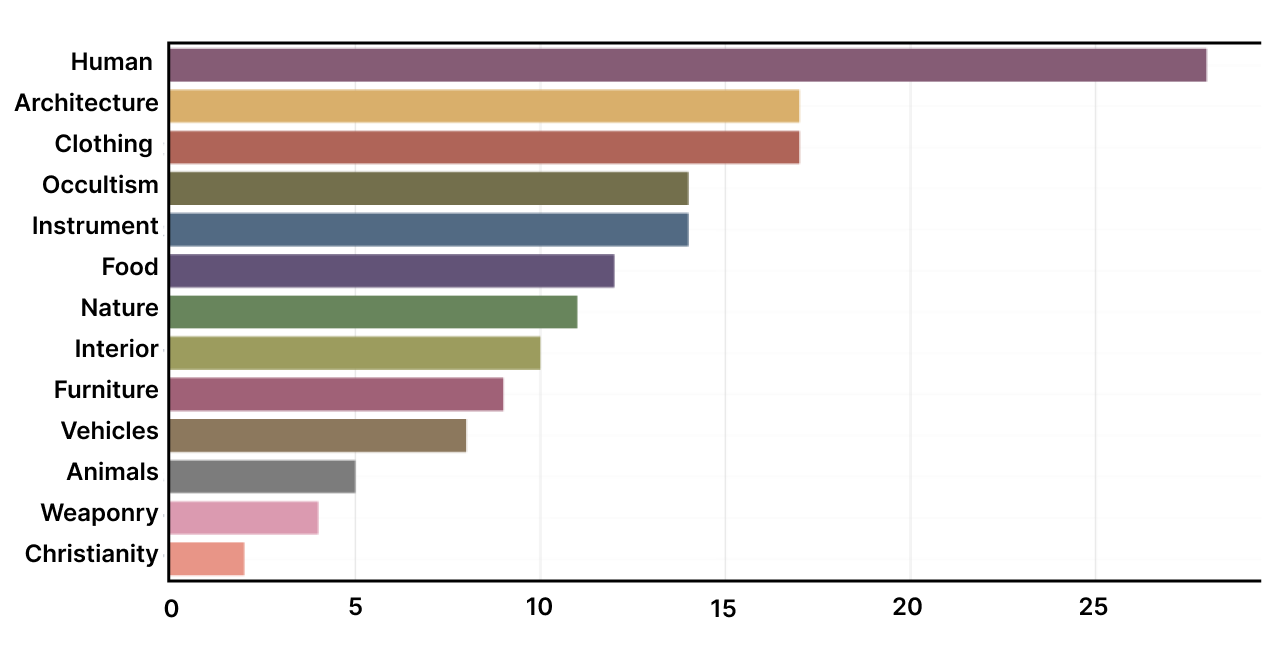}
  \caption{Total number of visits per category throughout the 22 sessions.}
  \Description{A horizontal bar chart displaying the total number of visits to each category. The chart visualizes the total visits per category in chronological order, with Human being the most visited category followed by Architecture, Clothing, Occultism, Instrument, Food, Nature, Interior, Furniture, Vehicles, Animals, Weaponry, and finally the least visited category, Christianity}
  \label{fig:categories-popularity}
\end{figure}

8 of the 22 participants mentioned the categories as an element that helped them follow their interests while exploring the collection. On average, each participant visited the Category Screen 6 times during their session. Throughout the 22 sessions all categories were visited, however the popularity of the categories varied, as illustrated in figure \ref{fig:categories-popularity}.

The log data revealed that 21 out of 22 participants explored the same category more than once during their session. In the interviews, multiple participants said they were drawn to unfold the content of the categories further, either going back and forth to it or by selecting filters in the category to narrow their search.
\begin{quote}
    \textit{``I chose human, I think, which is a bit broad, and then I went into it and I was like, I am interested to see women in the collection, so I started to select those to go deeper into a more narrow category.''} (P19)
\end{quote}

Similarly, the log data, showcasing which objects the participants saved during their session, supports the notion that participants wanted to investigate categories of interest more in-depth. The participants on average saved 6 objects during their session, and most participants (17) saved more than one object from the same category. One particularly eager participant saved 25 objects, all from the category ``Weaponry''. 

\subsection{Mislabelling and Interpretation}
\label{sec:mislabelling}
Out of the 22 test participants, 12 said that they noticed objects that were not correctly tagged. However, when asked if this influenced their experience none of the 12 participants said that it bothered them. Interestingly, participants seemed to express a large degree of understanding and perhaps even sympathy for the algorithm's mislabelling. Some suggested that it couldn't necessarily be determined what the correct label should be, pointing out that strict interpretation is not always possible. Others suggested that it was understandable why the object detection model would classify a given object as something other than it actually is because of its visual similarities, for instance a spear being labelled as a flute because of similar shape and colour:

\begin{quote}
    \textit{``Especially with the flute, he showed a lot of pictures of long, thin objects, which I do understand why he would think is a flute. And I always find this very interesting, because this is quite difficult, especially analyzing photos. It's quite difficult for artificial intelligence to do it. And as a human, you take a single look at it and you instantly know.''} (P12)
\end{quote}

Several of the participants that encountered incorrect labels found it interesting and said that being confronted with the AI’s “interpretation” made them question their own interpretation:

\begin{quote}
    \textit{``For example, for a mirror, there was one that was a full painting. That's why I clicked on it, I think, at some point, because I was like, that's not really a mirror. But I thought it was interesting because it kind of made you question whether it was you or the AI that was making a mistake, or it made you explore that.''} (P6)
\end{quote}

Several said that the incorrect labeling made them reflect or think differently about the potential visual interpretations of a particular object when taken out of context, thus challenging their own interpretation: 

\begin{quote}
    \textit{``I guess it made it a bit more exciting, because you didn't know if it was going to be the actual thing. One of them said it was a guitar, but it was open-heart surgery. It looked like a guitar. It was quite interesting.''} (P14)
\end{quote}

Interview participants often speculated why the AI had labeled an object the way it had. Particularly, participants noticed discrepancies in how the AI labelled the objects in contrast to how a human might interpret them. Similarly, people also speculated on what shared visual characteristics objects might have and how these shared characteristics would lead the AI to recognize a particular object incorrectly, but consistently: \textit{``I began to think about what the AI saw to think it was that object and what similarities it would have to the other objects''} (P10).

One participant stated that they thought the flaws were ``charming'' and that it made the AI seem more human. The same participant, however, also reflected on being misled and becoming suspicious of whether they could trust the AI at all when noticing a wrongly labeled object:

\begin{quote}
    \textit{``[...] all of a sudden, I became very aware that I suddenly couldn't trust it, that something I had clicked on and that it almost had me convinced was a skull, and I was like maybe it isn't that at all. That I am looking at it all of a sudden as an abstract, kind of distorted skull, but maybe it isn't.''} (P7)
\end{quote}
    
\subsection{(Re)contextualizing the Objects}
Second to the Object Screen, the participants spent the most time on the Canvas Screen (Table \ref{tab:avg-time-screens}), on which they could generate a new image using objects they had saved. 
When asked if playing with the objects on the canvas contributed to their interest in exploring the art or their experience of the artwork, most participants shared reflections concerning  (re)contextualization, composition in paintings, and piecing together different styles and details. 

\begin{quote}
    \textit{``I think it's just interesting to maybe take some details or take things in general, change the context and see what happens by reframing this relation.''} (P3)
\end{quote}

The opportunity to play with positioning objects on the canvas and creating new imagery made several participants contemplate how the objects were represented in the collection. 
\begin{quote}
    \textit{``Putting these objects together, you could give it your own context, and that also changed the way the objects were in the collection [...] it is an interesting way of combining images, not just generating completely new images, but combining specific objects from images to a completely new image was an interesting experience.''} (P10)
\end{quote}

\begin{figure*}[h]
  \centering
  \includegraphics[width=1\textwidth]{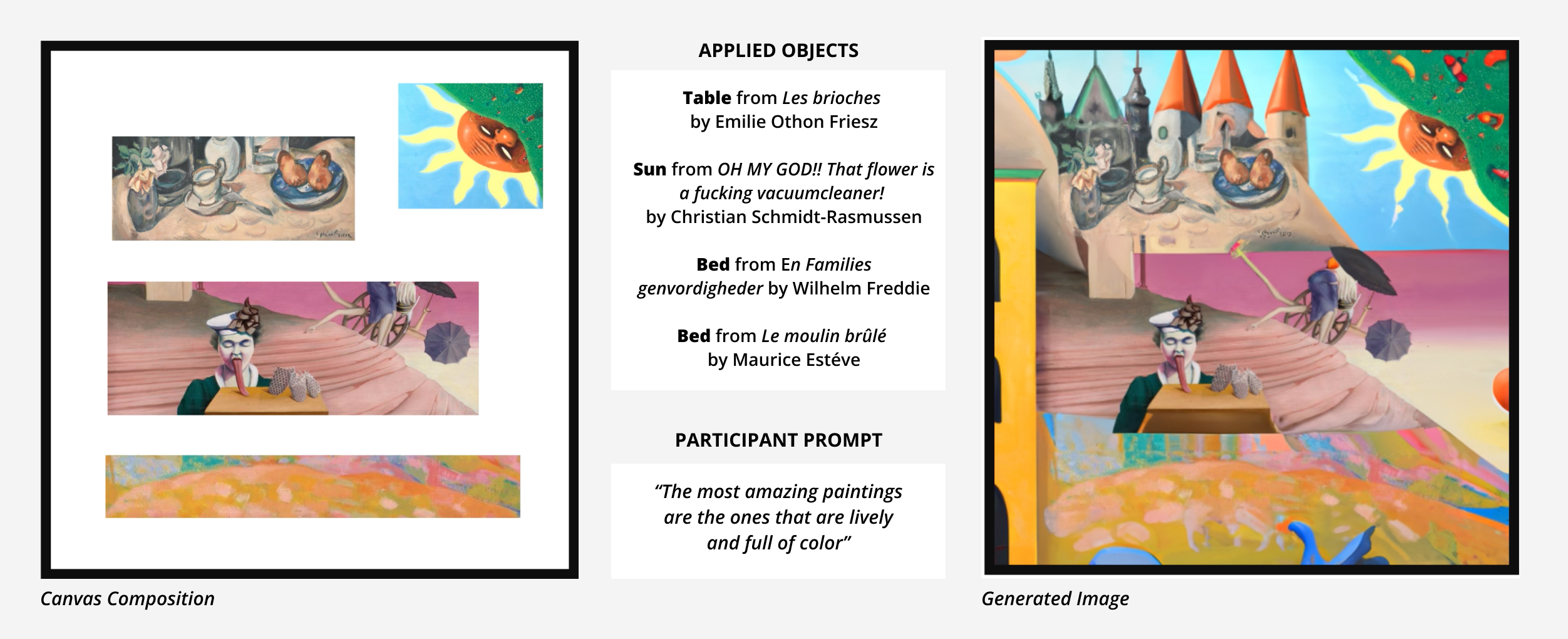}
  \caption{Canvas composition, applied objects, prompt and generated image by Participant 22.}
  \Description{A picture showcasing a square canvas before and after an image generation, alongside the text prompt provided by the user and a list of the objects used.}
  \label{fig:canvas-composition-example}
\end{figure*}

\begin{figure*}[h]
  \centering
  \includegraphics[width=1\textwidth]{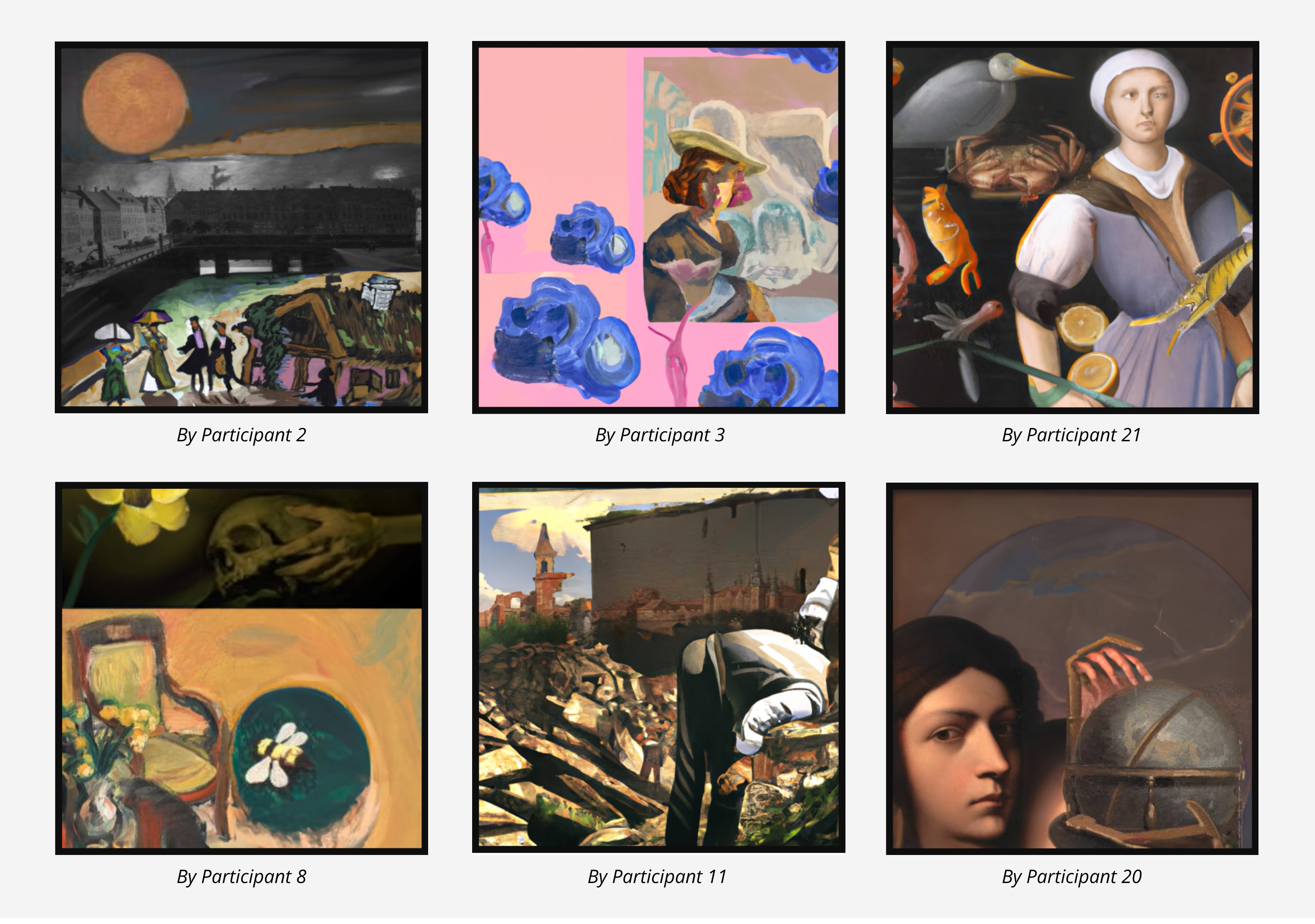}
  \caption{Example of images generated on the canvas.}
  \Description{A grid of 6 generated images by test participants, each different in level of detail and color scheme.}
  \label{fig:canvas-examples}
\end{figure*}

By placing the objects in a new and different context, the participants expressed that they became aware of compositional aspects of the art. This awareness concerned the composition of existing paintings and the composition they were creating on the canvas: \textit{``For instance, in the Baroque exhibition, there were a few areas in the paintings where there weren't a lot going on, and that didn't make sense to me. It made me think you could have added something''} (P8).

Furthermore, several participants also reflected on how combining objects from various parts of art history could reveal differences and similarities between styles across time: 
\begin{quote}
    \textit{``I thought it was cool how you could compose different pieces together. And I think if you did a bunch of different art movements together, you could learn a lot about how they evolved and how they could be intertwined.''} (P6)
\end{quote}

During testing, we noticed that the canvas increased the inclination of the participants to explore the collection further. All but one, spent a large enough amount of time exploring the various objects that they were asked to stop exploring and generate an image. When asked to do this, some asked for more time to find other objects. Others reflected on this aspect during the interviews, in which they expressed that had they been given more time, they would have gone back and collected more or different objects to use in their image, indicating a desire to investigate the collection even further.

When asked if anything unexpected happened while interacting with the application, the most frequent answer was that they were surprised by the resulting image they had generated on the canvas: 

\begin{quote}
    \textit{``I was really surprised to see an actual painting that could hang in a museum [...] The painting that AI generated reminded me really of one of my favorite artists. But none of the pictures were from him. So that's quite interesting. I really like that.''} (P22)
\end{quote}

Participants were in particular surprised by the way the outpainting functionality worked, saying that they did not expect the canvas to take the size and position of the object into account in the final result (although in fact most participants had opened an instruction screen which explained and visualized how the Canvas Screen worked). While several participants were familiar with other generative image systems such as Midjourney, participants were generally not familiar with outpainting.

\section{Discussion}
In the following we will reflect on four themes coming out of our design process as well as the testing and evaluation presented above: How the system affected the participants' view on the artworks, how the labelset influenced the design, the participants' experience of errors in the labelling, and the participants' experience with the Canvas Screen. Finally, we reflect on some implications for design.

\subsection{Experiencing Art Through the Lens of AI}
As proposed by Lev Manovich, machine learning offers new ways to experience art and visual culture by enabling the exploration of large collections and patterns, contrary to the traditional approach of inspecting artworks individually \cite{manovich_cultural_2020}. Through the evaluation we found that participants reflected on patterns in the museum collection, specifically objects recurring over multiple paintings. Seeing the recurrence of these objects side by side made them reflect on the motifs repeatedly depicted by artists and their various styles. In addition, the evaluation highlights that participants were inspired to focus more on details by exploring the collection through objects instead of full-sized paintings. 

It is particularly interesting to consider the participants' experience in light of the fact that they encountered our prototype after having visited the physical exhibitions at the museum. Many users expressed that they noticed new details and recurring objects in the art when exploring the application, which they had not noticed in the museum exhibition beforehand. Thus, the experience of exploring the art collection based on the objects detected by the machine learning model offered participants a new perspective on the art collection compared to the physical visit to the museum exhibition.  

\begin{figure}[h]
  \centering
    \includegraphics[width=0.90\linewidth]{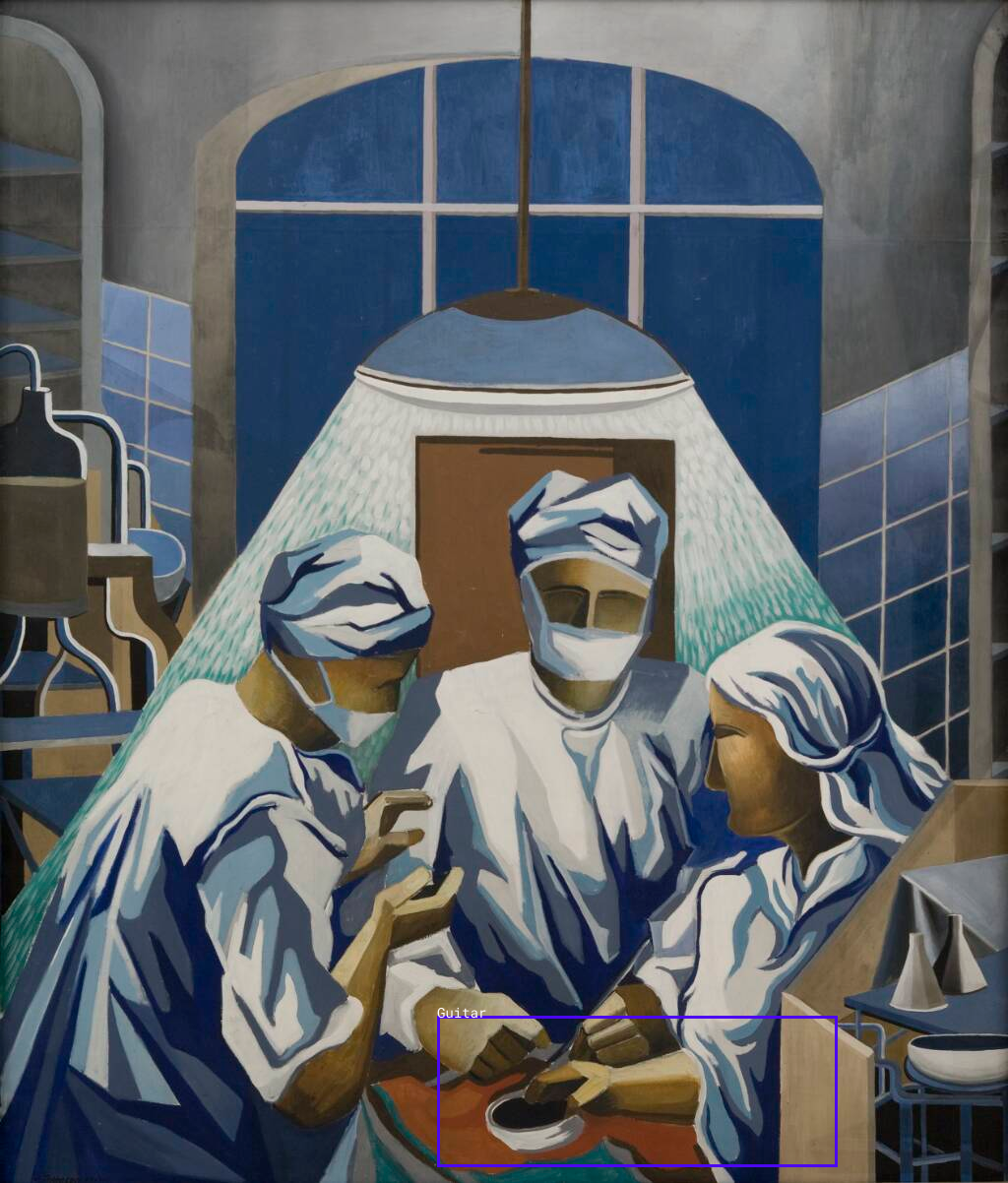}
    \caption{ \textit{Surgery} by J{\o}rgen Thomsen (1943-44). The detail in the middle lower part of the image was mislabelled as a ``guitar'' by GLIP.}
    \Description{Painting of three surgeons performing surgery and a bounding box around the object the model mistakenly classified as a guitar.}
    \label{fig:surgery}
\end{figure}

\subsection{Labelling}
The labels applied in an object detection model determine what objects can be detected - what the computer vision system can ``see'' in the image. The process of constructing the list of labels described above in section \ref{sec:ObjectDetection} demonstrates the importance of building a set of labels that enable the model to detect the most relevant objects. Our setup was limited by the amount of labels that could be fitted as an input string to the GLIP model - 120. This means that the model would not be able to label objects that were not included on the list in table \ref{tab:category-labels} - thus the objects labelled by the model represents only a partial view of all the objects in the collection, limited by the selection we had made. This may help explain the mislabelling of some details, such as that mentioned by participant 4 in section \ref{sec:mislabelling}, mistaking surgery for a guitar (see Figure \ref{fig:surgery}): Since the model did not include any labels relating to surgery, it could not label it correctly - instead settling on a label with some visual similarity but very different meaning. Indeed, in our first iteration of the object detection analysis using labels from COCO (see section \ref{sec:ObjectDetection}) this same detail was labeled 'bowl'; COCO does not have a label called 'guitar', nor any other labels that seem appropriate for this detail.

Given more time and technical resources, it might have been possible for us to increase the amount of different labels, by re-running GLIP over the artwork collection with different sets of labels each time. This might result in a dataset with several different labels for the same object, which would either have to be disambiguated through a separate process - or we could simply adjust the design to allow for multiple (possibly contradictory) labels for the same object, inviting users to reflect on the resulting ambiguity. We can only speculate on how such a larger set of labels would affect the user experience: One might hope that it would allow for an even richer experience with more nuance and more opportunities for surprising discoveries. However, it is also possible that adding more labels would increase the proportion of mislabelling, as the system would have to contend with a larger amount of categories overall while applying a limited ontology for each run of the object detection algorithm.

Future developments of GLIP and similar algorithms may lead to an increase in the number of labels that can be applied at a time. However, it is unlikely that this will remove all limitations on the ability of vision algorithms to detect objects in artwork. First, it may take some time before models can include a sufficiently large number of labels without forgoing precision: A comprehensive classification like Iconclass contains over 28,000 unique concepts, whereas the Getty Art \& Architecture Thesaurus contains 73,831 concept records, over 600 times larger than the number of labels used in our setup. Furthermore, even if a future system would allow a very long list of labels, there would remain some fundamental challenges with mapping between concepts and images precisely and comprehensively. Debates about the large image classification dataset ImageNet \cite{Russakovsky_imagenet_2015} have demonstrated that classifying images of humans with labels from a lexical database can lead to unintended consequences and controversy \cite{crawford_excavating_2019}. Similar complications may occur in object detection, as some concepts may mean different things at different times and cultural contexts. For instance, gender labels have acquired new meanings in recent times, adding nuance to what was formerly mostly considered a binary concept. Many other concepts relating to technology, societal roles, norms, institutions, or culture have changed meaning over time and in different societal and cultural contexts. Ciecko and colleagues \cite{ciecko_ai_2020} provide a striking example of how the use of image classification might inadvertently trigger controversy: An image of iron ankle manacles from Australia's convict history that is labeled as ``Fashion Accessory'' and ``Jewelry'' by a commonly used image classification algorithm. One could easily imagine that if a similar mislabelling were to occur in the collection of a museum relating to the history of slavery or the Holocaust, this could be offensive and hurtful for visitors and highly problematic for the museum.

In the first version of our labelset we included the label 'non-binary person', in order to accommodate a broader variation of gender identities and supplement the labels 'man' and 'woman'. However, the results made us question the classification. GLIP returned 210 bounding boxes with this label, of which the majority were depictions of children and/or nude people with displeased or uncomfortable facial expressions. We judged this to be potentially both inaccurate and harmful, and for these reasons we omitted this label from the final version of the labelset with the consequence that our application only provides two labels reflecting gender. This is unfortunate. While we do not have ground truth data available that could help us verify whether there are (few or many) images of people in the collection that should be tagged as non-binary person, the absence of this label might render invisible to the model a broader variety in gender identities. However, it seems that capturing nuances in gender presentation is difficult to do with the technology used in this study. It is worth reflecting on whether it is possible at all to classify gender with computer vision techniques that rely solely on visual appearance. For future work, it could be interesting to explore other ways to classify motifs of people in art instead of (or in addition to) gender, e.g. by clothing, hair, age etc.

\subsection{Mislabelling and Trust}
Seen in light of the challenges with labelling objects correctly outlined above, it is striking that the test participants generally trust our system's algorithmic labeling. Only 12 of the 22 participants noticed objects that were incorrectly labeled, even though mislabeled objects -- or at least questionably labeled -- can easily be found in most categories. (Consider, for instance, some of the objects shown in Fig. \ref{fig:teaser} and \ref{fig:cropped-images}.) Those who did notice questionable labels often seemed willing to offer explanations on behalf of the algorithm, one participant even personifying the algorithm: \textit{``...he showed a lot of pictures of long, thin objects, which I do understand why he would think is a flute''} (P12). Others suggested the mislabelling made them question their own interpretations - which is well aligned with the typical ideals of art education in museums, which often emphasize questioning one's preconceptions and interpretation and opening oneself up to seeing artworks in different ways.

While these observations align with other research pointing towards a tendency to overtrust in AI systems \cite{howard2020we, robinette2016overtrust,benford_sensitive_2022}, we do not have data to assess clearly why the participants were so willing to trust the system or make excuses for its errors. However, there is a striking similarity with the observations done by Benford and colleagues when exploring the use of emotion detection AI in an art museum: ``...visitors tended to construct post hoc rationalizations of their emotional experience that agreed with, or at least accommodated, the 'results' reported by the system, even when this differed from their initial reflections'' \cite[p.12]{benford_sensitive_2022}.

We can only speculate about why visitors appear so willing to trust in the output of these computer vision systems -- object detection in our case, emotion detection in \cite{benford_sensitive_2022}. First, several factors in the presentation at the museum may inspire trust among visitors: The system is presented to them by university researchers, which may influence the participants to see the system as trustworthy and authoritative; and the context of the museum as a highly trusted institution may add to this impression. Second, visitors may be extra understanding towards the system's errors due to the application domain, as interpreting art is both a difficult task and often seen not to have a single correct answer and one to which computer systems are not commonly applied. Third, given the large amount of visual information in the interface and the focus on exploration, it is possible that some mislabellings - like those showing unclear images and shapes - were overlooked as ``noise'' as participants focused on the higher resolution, and thus, more clear and recognizable images.

\begin{figure}[h]
  \centering
    \includegraphics[width=0.90\linewidth]{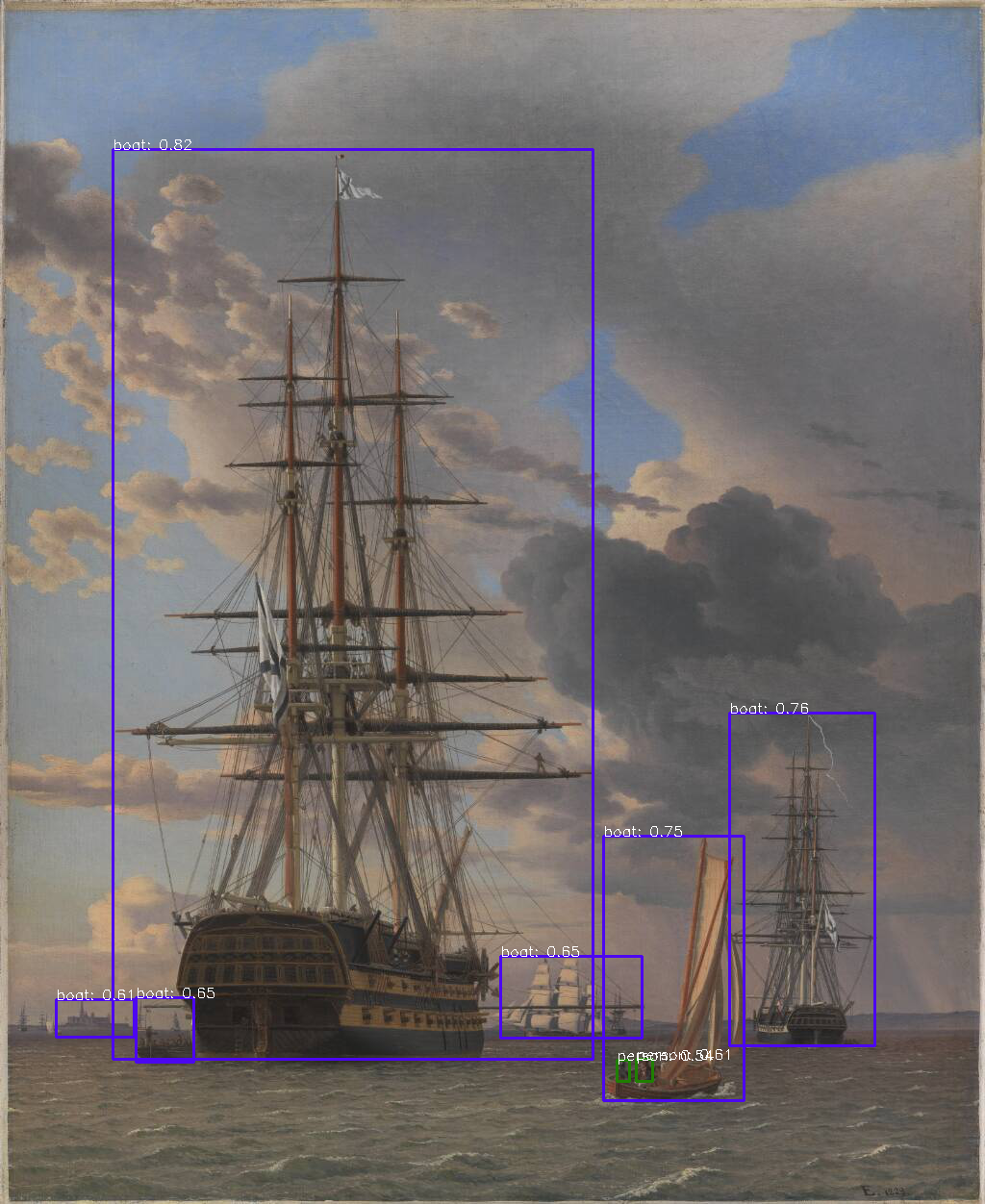}
    \caption{Classification of objects is sometimes affected by the context. In this image, five sea-faring vessels of varying sizes are correctly classified as 'boat'. However the small, left-most blue bounding box surrounds the outline of the castle Kronborg which is wrongly classified as 'boat' - presumably due to the vicinity to the sea and other boats. Artwork: \textit{The Russian Ship of the Line "Asow" and a Frigate at Anchor near Elsinore} by C.W. Eckersberg (1828).}
    \Description{Painting of sea-faring vessels on the ocean and several bounding boxes showcasing the objects the model has detected.}
    \label{fig:ships}
\end{figure}

To some degree, the discussion above has presupposed that there is a correct and an incorrect label for each object in an art image. That assumption might be challenged in several ways. First, art images frequently appear ambiguous and resist interpretation from audiences, art critics and scholars alike. For instance, should Salvador Dalí's ``Lincoln in Dalivision'' be classified as depicting the face of a bearded man, or a naked woman standing by a window? Much art is even more abstract and difficult to interpret unambiguously; and as ambiguity is a central quality of art, removing ambiguity from art is not a desirable goal. Second, one might argue that computer vision may encode ways of seeing objects that subtly differ from our assumptions about how objects should be seen - and which may offer interesting perspectives. For instance, Leahu demonstrates that a vision algorithm may end up encoding not just objects as discrete entities, but also aspects of the \textit{relations} that constitute them - such as when a neural network trained to recognize dumbbells also encodes the images of arms holding the dumbbells \cite{leahu_ontological}. For Leahu, this raises the possibility of "ontological surprises" - that computer vision algorithms may reveal unexpected relations between objects. In our analysis with GLIP we could sometimes see that the context surrounding an object might affect the algorithm's classification of objects, as in the example in Figure \ref{fig:ships}. It would be an interesting challenge for future research to explore whether this sensitivity to context - or other particular aspects of the way computer vision encodes objects - could be used to help art viewers or even art scholars discover new ways of seeing art.

\subsection{Creating New Images}
As highlighted in Section \ref{design-section}, we included the canvas feature to support user engagement in exploration of the collection. Through the test, we found that the canvas encouraged participants to continue their search for objects: When given the task of using the Canvas Screen to make an image, many users were eager to go back and look for more objects they could use to make images. Several also said they would have liked to spend more time going back and forth between the Canvas Screen and the collection. One particularly eager user (P8) spent a long while creating multiple images and would only stop when we insisted that we needed to end the testing session. These observations confirm that the generative feature helped support engagement.

In addition, we found that generating an image through positioning and combining objects on the canvas made the participants reflect on the artworks' context, time periods, styles, details, and composition. With outpainting, the participants were able to visually experience how styles and details can be merged into something new that goes beyond the original context of the object(s) (Fig. \ref{fig:canvas-composition-example}). This suggests that outpainting may have a promising potential as a device to facilitate practice-based learning about these dimensions of visual art in a manner that would be much more rapid and less dependent on practical skills than traditional exercises in drawing and painting. 

\subsection{Implications for Design}
Based on the observations outlined above, we suggest a few topics that might be relevant to consider for designers working with object detection in digital art collections.

First, designers should pay close attention to the labelset used for object detection. As long as object metadata for the collection is unavailable or incomplete, it will be difficult to assess - other than by trial and error - which types of objects are prevalent in the collection and can be detected reliably. However, working with subject experts like museum curators or art historians might help identify appropriate labels, particularly when working with collections dominated by older art.

Second, designers might be interested in deliberately introducing flaws or errors in labeling as a way to provoke reflection and nudge users to question their own interpretation. However, our observations suggest that such errors might need to stand out strongly to ensure that users will notice them and identify them as errors. If users place trust and even some sympathy with the algorithm, then designers who wish to inspire critical reflection about the algorithm will need to work carefully on communicating to users that the algorithm is not necessarily to be trusted fully. Designers might explore ways to include confidence measures or other ways of visualizing uncertainty in labelling, however this would need to be balanced against the need to avoid disrupting the aesthetic of the art presentation, which is a strong design norm in art museums. Alternatively, designers might create deliberately ambiguous presentations of the algorithm's outputs in order to provoke critical reflection, following tactics similar to those presented in \cite{ambiguity, staying_open}.

Third, future designers and art educators might use generative systems (such as in our Canvas Screen) to facilitate learning about visual art and composition. For instance, one might use a more narrowly curated set of paintings based on time period or style to provide insight into frequently depicted motifs and typical composition. This could be supported by predefined text prompts that exemplify styles, details, and compositions recurring within the particular collection of artworks, allowing the user to explore objects and visually experiment with image generation while working towards more focused learning outcomes. 

\section{Conclusion}
We have presented an approach to using object detection to facilitate exploration of a large digital art collection. First, we have demonstrated that recent leaps in computer vision, in particular the emergence of multimodal models like CLIP and GLIP, has made it feasible to use object detection on digitised art images with sufficient precision to support a meaningful and satisfying user experience for a general art-interested audience such as the visitors to the National Gallery of Denmark.

Second, we have presented the design of a web application that use the object detection data as basis for an interface that allow users to explore the collection in a novel way, using objects of interest as an entry point, and using a generative system with outpainting to facilitate creative and playful exploration. The evaluation has demonstrated that this interface has inspired test participants to see the art in a new light and discover new things about the art. We have highlighted the importance of constructing an appropriate labelset for the object detection, and drawn awareness to the participants' tendency to trust the system's output and perhaps overlook errors in the object labelling. Finally, we have suggested some design implications that might inform future work with object detection in artwork.

Our study has been limited to only artworks classified as paintings in the museum collection. Further research would be needed to explore whether the technology can be applied across diverse media types such as drawings and sketches, sculptures, photos and video, engravings, and so on. Furthermore, there is a need for cross-disciplinary research collaboration with art experts (for instance in art history or the digital humanities) to explore the aesthetic and pedagogical implications of extracting details from their original context in the artworks and presenting them to users as lists of objects from a variety of different artworks, styles, periods and artistic agendas. While such an approach may seem problematic for some curators as it means that image fragments are presented detached from their original context in the artwork, our study has demonstrated that it has the potential to inspire and engage museum visitors to discover and learn more about art. Tapping into this potential would be beneficial for both museums and their visitors - and would break new ground for the use of computer vision in art education and dissemination.

\begin{acks}
This work was supported by a research grant (40575) from VILLUM FONDEN. We thank Jonas Heide Smith and Nikolaj Erichsen at the National Gallery of Denmark for their help and support. 
\end{acks}

\bibliographystyle{ACM-Reference-Format}
\bibliography{SMKExplore}

\end{document}